# Three Applications of the String-Inspired Technique to Quantum Electrodynamics *

D. Fliegner [a], M. G. Schmidt [a], C. Schubert [b] [†]

[a] *Institut für Theoretische Physik, Universität Heidelberg*
*Philosophenweg 16, D-69120 Heidelberg, Germany*

[b] *Humboldt Universität zu Berlin*
*Invalidenstr. 110, D-10115 Berlin, Germany*

## Abstract

We discuss the following recent applications of the "string-inspired" worldline technique to calculations in quantum electrodynamics:
i) Photon splitting in a constant magnetic field, ii) The two-loop Euler-Heisenberg Lagrangian, iii) A progress report on a recalculation of the three-loop QED $\beta$ – function.

---

*Talk given by C. Schubert at the DESY Workshop on QCD and QED in higher orders, Rheinsberg, Germany, 21st-26th April, 1996.
†Supported by Deutsche Forschungsgemeinschaft.

# 1. Introduction

Since the pioneering work of Bern and Kosower [1] it is known that techniques from string perturbation theory can be used to improve on calculational efficiency in ordinary quantum field theory. Those authors obtained a novel type of parameter integral representations for one-loop amplitudes in ordinary quantum field theory by representing them as the infinite string tension limits of appropriately chosen (super) string amplitudes. By a detailed analysis of this limit, the construction of these integral representations was summarized in terms of the "Bern-Kosower Master Formula", and the "Bern-Kosower Rules". Those rules do not explicitly refer to string theory any more, and have been found useful for both gluon [2] and graviton scattering [3] calculations.

An alternative and more elementary approach to this subject was later developed by Strassler [4], based on the known representations of quantum field theoretical amplitudes in terms of (super) particle path integrals [5, 6]. For many cases of interest, he was able to show that an evaluation of those path integrals in analogy to string perturbation theory leads to parameter integrals which are identical to those encoded in the corresponding Bern-Kosower master formula. Due to its simplicity, this approach to the Bern-Kosower formalism has turned out to be suitable for multiloop generalizations [7, 8, 9], to the inclusion of constant background fields [10, 11], and the combination of both [12].

In this report, we concentrate on the case of photon scattering in spinor QED. We will first shortly review Strassler's formalism for this special case, and then discuss three recent or ongoing calculations performed with generalizations of this technique: i) One-loop photon splitting in a constant magnetic field [13]; ii) The two-loop correction to the Euler-Heisenberg Lagrangian [12]; iii) A recalculation of the three-loop QED $\beta$ – function [14].

# 2. One–Loop Photon Scattering in the Worldline Formalism

In his rederivation of the Bern–Kosower rules for photon scattering off a spinor loop, Strassler sets out from the following well-known path integral representation for the corresponding one–loop effective action (see e.g. [6]):

$$\Gamma[A] = -2\int_0^\infty \frac{dT}{T} e^{-m^2 T} \int \mathcal{D}x \mathcal{D}\psi \exp\left[-\int_0^T d\tau \left(\frac{1}{4}\dot{x}^2 + \frac{1}{2}\psi\dot{\psi} + ieA_\mu \dot{x}^\mu - ie\psi^\mu F_{\mu\nu}\psi^\nu\right)\right] \quad (1)$$

Here $T$ is the usual Schwinger proper–time parameter, the $x^\mu(\tau)$'s are the periodic functions from the circle with circumference $T$ into $D$ – dimensional spacetime, and the $\psi^\mu(\tau)$'s their antiperiodic Grassmannian supersymmetric partners.

In the "string–inspired" approach, the path integrals over $y$ and $\psi$ are evaluated by one-dimensional perturbation theory, using the Green functions

$$\begin{aligned}
\langle y^\mu(\tau_1) y^\nu(\tau_2)\rangle &= -g^{\mu\nu} G_B(\tau_1, \tau_2) \\
&= -g^{\mu\nu}\left[|\tau_1 - \tau_2| - \frac{(\tau_1 - \tau_2)^2}{T}\right] \\
\langle \psi^\mu(\tau_1)\psi^\nu(\tau_2)\rangle &= \frac{1}{2} g^{\mu\nu} G_F(\tau_1, \tau_2) \\
&= \frac{1}{2} g^{\mu\nu} \text{sign}(\tau_1 - \tau_2)
\end{aligned} \quad (2)$$

We will often abbreviate $G_B(\tau_1, \tau_2) =: G_{B12}$ etc. The bosonic Wick contraction is actually carried out in the relative coordinate $y(\tau) = x(\tau) - x_0$ of the closed loop, while the (ordinary) integration over the average position

$$x_0 = \frac{1}{T}\int_0^T d\tau\, x(\tau)$$

yields energy–momentum conservation. With our conventions, the free bosonic path integral is normalized as

$$\int \mathcal{D}y \exp\left[-\int_0^T d\tau \frac{1}{4}\dot{y}^2\right] = [4\pi T]^{-\frac{D}{2}},$$

while the free fermionic one is equal to unity. The result of this evaluation is the one-loop effective Lagrangian $\mathcal{L}(x_0)$.

One–loop scattering amplitudes are obtained by specializing the background to a finite sum of plane waves of definite polarization. Equivalently, one may define integrated vertex operators

$$V = \int_0^T d\tau \left[\dot{x}^\mu \varepsilon_\mu - 2\mathrm{i}\psi^\mu\psi^\nu k_\mu \varepsilon_\nu\right] \exp[ikx(\tau)] \tag{3}$$

for external photons of definite momentum and polarization, and calculate multiple insertions of those vertex operators into the free path integral.

Alternatively, the $x$- and $\psi$- path integrals may be combined into a super path integral [6]

$$\Gamma[A] = -2\int_0^\infty \frac{dT}{T} e^{-m^2 T} \int \mathcal{D}X \exp\left[-\int_0^T d\tau \int d\theta \left(-\frac{1}{4}X D^3 X - \mathrm{i}e D X^\mu A_\mu(X)\right)\right], \tag{4}$$

with the definitions

$$X^\mu = x^\mu + \sqrt{2}\,\theta\psi^\mu = x_0^\mu + Y^\mu,\ D = \frac{\partial}{\partial\theta} - \theta\frac{\partial}{\partial\tau}.$$

The photon vertex operator is then rewritten as

$$V = -\int_0^T d\tau\, d\theta\, \varepsilon_\mu\, DX^\mu \exp[ikX],$$

and we are left with a single Wick–contraction rule

$$\begin{aligned}
\langle Y^\mu(\tau_1,\theta_1) Y^\nu(\tau_2,\theta_2)\rangle &= -g^{\mu\nu}\hat{G}(\tau_1,\theta_1;\tau_2,\theta_2),\\
\hat{G}(\tau_1,\theta_1;\tau_2,\theta_2) &= G_B(\tau_1,\tau_2) + \theta_1\theta_2 G_F(\tau_1,\tau_2).
\end{aligned} \tag{5}$$

### 3. Photon Splitting in a Constant Magnetic Field

Photon splitting in a strong magnetic field is a process of potential astrophysical interest. The first calculation valid for arbitrary photon frequency $\omega$ below the pair creation threshold was given by Adler [15], who obtained the exact amplitude as a strongly convergent triple integral. Recently, renewed interest in this subject was triggered by a new calculation of this quantity due to Mentzel, Berg and Wunner [16], and a subsequent claim by Wunner, Sang and Berg [17] that numerical evaluation of the resulting formula yields photon splitting rates roughly four orders of magnitude larger than those found in [15] (it has now been found that this claim was based on a computer code error [18]).

This has prompted us [13] to use the worldline formalism for the following recalculation of the one-loop three-photon amplitude in a constant magnetic field.

As in field theory, the constant magnetic background field $B$ is best taken into account in Fock–Schwinger gauge, where $A_\mu = \frac{1}{2} x^\rho F_{\rho\mu}$. Its contribution to the worldline Lagrangian then becomes

$$\Delta \mathcal{L} = \frac{1}{2}\, ie\, y^\mu F_{\mu\nu} \dot{y}^\nu - ie\, \psi^\mu F_{\mu\nu} \psi^\nu \quad .$$

Being bilinear, those terms can be simply absorbed into the kinetic part of the Lagrangian. This leads to generalized worldline propagators ( [11]; see also [19]) which we write in the form [12]

$$\mathcal{G}_B(\tau_1, \tau_2) = \frac{1}{2(eF)^2} \left( \frac{eF}{\sin(eTF)} e^{-ieTF\dot{G}_{B12}} + ie\, F \dot{G}_{B12} - \frac{1}{T} \right)$$

$$\mathcal{G}_F(\tau_1, \tau_2) = G_{F12} \frac{e^{-ieTF\dot{G}_{B12}}}{\cos(eTF)} \tag{6}$$

(a dot always denotes a derivative with respect to the first variable). Those now have non-vanishing coincidence limits, which for $\mathcal{G}_B$ can be subtracted due to momentum conservation. Again $\mathcal{G}_B$ and $\mathcal{G}_F$ may be assembled into a superpropagator,

$$\hat{\mathcal{G}}(\tau_1, \theta_1; \tau_2, \theta_2) \equiv \mathcal{G}_B(\tau_1, \tau_2) + \theta_1 \theta_2 \mathcal{G}_F(\tau_1, \tau_2) \quad .$$

To obtain the photon splitting amplitude, we will use this Green function for the Wick contraction of three vertex operators $V_0$ and $V_{1,2}$, representing the incoming and the two outgoing photons.

The calculation is greatly simplified by the special kinematics of this process. Energy-momentum conservation $k_0 + k_1 + k_2 = 0$ forces collinearity of all three four-momenta, so that, writing $-k_0 \equiv k \equiv \omega n$,

$$k_1 = \frac{\omega_1}{\omega} k,\, k_2 = \frac{\omega_2}{\omega} k;\quad k^2 = k_1^2 = k_2^2 = k \cdot k_1 = k \cdot k_2 = k_1 \cdot k_2 = 0. \tag{7}$$

Moreover, a simple CP-invariance argument together with an analysis of dispersive effects [15] shows that there is only one allowed polarization case. This is the one where the incoming photon is polarized parallel to the plane containing the external field and the direction of propagation, and both outgoing ones are polarized perpendicular to this plane. This choice of polarizations leads to the further vanishing relations

$$\varepsilon_{1,2} \cdot \varepsilon_0 = \varepsilon_{1,2} \cdot k = \varepsilon_{1,2} \cdot F = 0 \quad . \tag{8}$$

This leaves us with the following small number of nonvanishing Wick contractions:

$$\langle V_0 V_1 V_2 \rangle = -i \prod_{i=0}^{2} \int_0^T d\tau_i \int d\theta_i\, \exp\left[\frac{1}{2}\sum_{i,j=0}^{2} \bar{\omega}_i \bar{\omega}_j\, n \hat{\mathcal{G}}_{ij} n\right] \varepsilon_1 D_1 D_2 \hat{\mathcal{G}}_{12} \varepsilon_2 \sum_{i=0}^{2} \bar{\omega}_i \varepsilon_0 D_0 \hat{\mathcal{G}}_{0i} n \tag{9}$$

For compact notation we have defined $\bar{\omega}_0 = \omega,\, \bar{\omega}_{1,2} = -\omega_{1,2}$. This result must still be multiplied by an overall factor of

$$\frac{(eTB)\cosh(eTB)}{(4\pi T)^2 \sinh(eTB)} \quad ,$$

to take into account the change of the path integral determinant due to the external field [10].

It is then a matter of simple algebra to obtain the following representation for the matrix element $C_2[\omega,\omega_1,\omega_2,B]$ appearing in eq. (25) of [15]:

$$C_2[\omega,\omega_1,\omega_2,B] = \frac{m^8}{4\omega\omega_1\omega_2}\int_0^\infty dT\, T\frac{e^{-m^2T}}{z^2\sinh(z)}$$

$$\int_0^T d\tau_1\, d\tau_2\, \exp\left\{-\frac{1}{2}\sum_{i,j=0}^2 \bar{\omega}_i\bar{\omega}_j\left[G_{Bij} + \frac{T}{2z}\frac{\cosh(z\dot{G}_{Bij})}{\sinh(z)}\right]\right\}$$

$$\times\left\{\left[-\cosh(z)\ddot{G}_{B12}+\omega_1\omega_2\Big(\cosh(z)-\cosh(z\dot{G}_{B12})\Big)\right]\right.$$

$$\times\left[\frac{\omega}{\sinh(z)\cosh(z)}-\omega_1\frac{\cosh(z\dot{G}_{B01})}{\sinh(z)}-\omega_2\frac{\cosh(z\dot{G}_{B02})}{\sinh(z)}\right]$$

$$\left.+\frac{\omega\omega_1\omega_2 G_{F12}}{\cosh(z)}\left[\sinh(z\dot{G}_{B01})\Big(\cosh(z)-\cosh(z\dot{G}_{B02})\Big)-(1\leftrightarrow 2)\right]\right\}. \quad (10)$$

Translation invariance in $\tau$ has been used to set the position $\tau_0$ of the incoming photon equal to $T$, and we have abbreviated $eTB \equiv z$.

Numerical analysis of eq. (10) has shown it to be in complete agreement with its counterpart in [15], as well as with the formulas by Stoneham [20] and the recent result of Baier et al. [21]. A package of FORTRAN programs for the numerical evaluation of those various formulas may be obtained at S.L. Adler's home page (http://www.sns.ias.edu/~adler/Html/photonsplit.html).

**4. The Two-Loop Euler-Heisenberg Lagrangian**

This calculus may be generalized to the two-loop level in the following way [12]. As in the case without an external field, we can represent an internal photon correction in the fermion loop by inserting, into the path integral representing the fermion loop, the following worldline current-current interaction term [8],

$$\frac{e^2}{2}\frac{\Gamma(\lambda)}{4\pi^{\lambda+1}}\int_0^T d\tau_a\, d\theta_a\int_0^T d\tau_b\, d\theta_b\, \frac{DX_a^\mu DX_{b\mu}}{\left((X_a-X_b)^2\right)^\lambda} \quad (11)$$

($\lambda = \frac{D}{2}-1$). The denominator of this insertion is written in Schwinger proper-time representation and becomes

$$\int_0^\infty d\bar{T}(4\pi\bar{T})^{-\frac{D}{2}}\exp\left[-\frac{(X(\tau_a)-X(\tau_b))^2}{4\bar{T}}\right].$$

For fixed proper-time $\bar{T}$, the exponent may be regarded as another additional term in the worldline Lagrangian for the loop path integral. Being quadratic in the superfield $X$, it can again be absorbed into the worldline superpropagator, leading to the following Green function,

$$\hat{\mathcal{G}}^{(1)}(\tau_1,\tau_2) = \hat{\mathcal{G}}(\tau_1,\tau_2) + \frac{1}{2}\frac{[\hat{\mathcal{G}}(\tau_1,\tau_a)-\hat{\mathcal{G}}(\tau_1,\tau_b)][\hat{\mathcal{G}}(\tau_a,\tau_2)-\hat{\mathcal{G}}(\tau_b,\tau_2)]}{\bar{T}-\frac{1}{2}\hat{\mathcal{C}}_{ab}}, \quad (12)$$

$\hat{\mathcal{C}}_{ab} \equiv \hat{\mathcal{G}}(\tau_a,\tau_a) - \hat{\mathcal{G}}(\tau_a,\tau_b) - \hat{\mathcal{G}}(\tau_b,\tau_a) + \hat{\mathcal{G}}(\tau_b,\tau_b)$. This object carries the full information on the ("scalar part" of) the inserted photon propagator, as well as on the external field, and the electron spin. It replaces the one-loop Green function Green function $\hat{\mathcal{G}}$ for two-loop calculations in an external field, only that now we have to include into the Wick contractions the "left-over" numerator $D_a y_a^\lambda D_b y_{b\lambda}$ of the photon insertion, and there are three more parameter integrations to perform. Considering just the simplest case, which is the two-loop correction to the Euler-Heisenberg Lagrangian, this allows us to write down without further ado the following integral representation for this Lagrangian,

$$\mathcal{L}^{(2)}_{spin}[F] = \frac{e^2}{(4\pi)^D} \int_0^\infty dT\, d\bar{T}\, \frac{\mathrm{e}^{-m^2 T}}{T^{1+\frac{D}{2}}} \det{}^{-\frac{1}{2}}\left[\frac{\tan(eFT)}{eFT}\right] \int_0^T d\tau_a d\tau_b \int d\theta_a d\theta_b$$

$$\times \det{}^{-\frac{1}{2}}\left[\bar{T} - \frac{1}{2}\hat{\mathcal{C}}_{ab}\right] \mathrm{tr}\left[D_a D_b \hat{\mathcal{G}}_{ab} + \frac{1}{2}\frac{D_a(\hat{\mathcal{G}}_{aa}-\hat{\mathcal{G}}_{ab})D_b(\hat{\mathcal{G}}_{ab}-\hat{\mathcal{G}}_{bb})}{\bar{T}-\frac{1}{2}\hat{\mathcal{C}}_{ab}}\right], \qquad (13)$$

where the notations should be obvious. The further evaluation and renormalization of this Lagrangian, as well as a comparison with earlier calculations (see [22] and references therein) may be found in [12].

### 5. The Three-Loop QED $\beta$ – Function

Finally, let us remark on work in progress on a recalculation of the 3-loop QED $\beta$ – function [14]. While this $\beta$ – function coefficient has been calculated many years ago [23], that calculation and all subsequent ones have been plagued by extensive cancelations between diagrams. In particular, the "quenched" contribution to this function due to vacuum polarization diagrams with a single spinor loop and two photon insertions, which is scheme-independent, turns out to have a rational coefficient, despite of the appearance of terms involving $\zeta(3)$ for individual diagrams (for a recent knot-theoretical explanation of this fact see [24]). Even more dramatic cancelations occur if this calculation is done in Feynman gauge and Pauli-Villars regularization [25].
These findings clearly make it an interesting application for the present formalism, which avoids the split of the vacuum polarization amplitude into individual diagrams, as explained in [7, 8]. The calculation of the quenched QED three-loop $\beta$ – function which we are currently undertaking is a straightforward generalization of the two-loop calculation presented in [8]. Our approach is to calculate the three-loop vacuum amplitude in a constant external field $F$, using the three-loop generalization $\hat{\mathcal{G}}^{(2)}$ [8] of the two-loop worldline super Green function eq. (12), and then to extract the induced Maxwell term $\sim \frac{1}{\epsilon}F^{\mu\nu}F_{\mu\nu}$.
While this calculation is not yet finished, we would like to present here the method we employ for calculating the arising parameter integrals.
The most basic integral appearing in this calculation reads, in dimensional regularization,

$$I(D) = \int_0^\infty dT_1\, dT_2 \int_0^1 da\, db\, dc\, dd \left[(T_1 + G_{Bab})(T_2 + G_{Bcd}) - \frac{C^2}{4}\right]^{-\frac{D}{2}} \qquad (14)$$

Here $T_1$, $T_2$ denote the proper-time lengths of the inserted photon propagators, $C \equiv G_{Bac} - G_{Bad} - G_{Bbc} + G_{Bbd}$, and the $G_{Bij}$ are the ordinary one-loop Green functions from eq. (2). The electron proper-time integral decouples in the $\beta$ – function calculation, and just yields an overall factor of

$$\int_0^\infty \frac{dT}{T} \mathrm{e}^{-m^2 T} T^{6-\frac{3}{2}D} = \Gamma(6 - \frac{3}{2}D)\, m^{3D-12} \quad.$$

The nontrivial integrations are $\int_0^1 da\, db\, dc\, dd \equiv \int_{abcd}$, representing the four end points of the photon propagators moving around the electron loop (fig.1). The integrand has one trivial invariance under the operator $\frac{\partial}{\partial a} + \frac{\partial}{\partial b} + \frac{\partial}{\partial c} + \frac{\partial}{\partial d}$, which just shifts the location of the zero on the loop. The integral decomposes into 24 ordered sectors, of which 16 constitute the planar (P) (fig. 1a) and 8 the nonplanar (NP) sector (fig. 1b). Due to the symmetry properties of the integrand, all sectors of the same topology give an equal contribution.

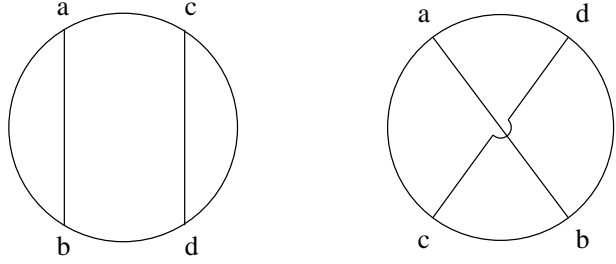

Figure 1:     1a                 1b

A general integral encountered in this calculation has the same denominator, possibly with a different power, and a numerator, which is again expressible in terms of the functions $G_{Bab}$, $G_{Bcd}$, $C$ and their derivatives.

As a first step in the calculation of $I(D)$, we add and subtract the same integral with $C = 0$ and rewrite $I(D) = I_{sing}(D) + I_{reg}(D)$,

$$I_{sing}(D) = \int_0^\infty dT_1\, dT_2 \int_{abcd} \left[(T_1 + G_{Bab})(T_2 + G_{Bcd})\right]^{-\frac{D}{2}} \qquad (15)$$

$I_{sing}(D)$ can be calculated right away, as it factorizes into two elementary three-parameter integrals, yielding

$$I_{sing}(D) = \left[\frac{2B(2 - \frac{D}{2}, 2 - \frac{D}{2})}{D - 2}\right]^2$$

Inspection of $I_{reg}(D)$ reveals that it contains no divergences, so that we can set $D = 4$ in its calculation (it still contributes to the $\frac{1}{\epsilon}$ - pole due to the global $\Gamma(6 - \frac{3}{2}D)$). The integrations over $T_1, T_2$ are then elementary, and we are left with

$$I_{reg}(4) = \int_{abcd} \left[-\frac{4}{C^2}\ln\left(1 - \frac{C^2}{4G_{Bab}G_{Bcd}}\right) - \frac{1}{G_{Bab}G_{Bcd}}\right] \qquad (16)$$

To calculate this integral, we will project it onto the nonplanar and planar sectors in turn, $I_{reg}(4) = I_P + I_{NP}$. We define $D_{ab} \equiv \frac{\partial}{\partial \tau_a} + \frac{\partial}{\partial \tau_b}$, and note the identities

$$D_{ab}C = \pm 2\chi_{NP},\ D_{ab}^2 C = 2\delta_{abcd}\ , \qquad (17)$$

where $\delta_{abcd} \equiv \delta_{ac} - \delta_{ad} - \delta_{bc} + \delta_{bd}$, and $\chi_{NP}$ denotes the characteristic function of the nonplanar sector. Given that $D_{ab}$ does not act on $G_{Bab}$ and $G_{Bcd}$, this implies a second invariance of the integrand valid only in the planar sector; in this sector, we can even keep one propagator fixed, and just shift the other one. From these identities, the trivial identity

$$\int_{abcd} D_{ab}^2\, g(G_{Bab}, G_{Bcd}, C) = 0$$

valid for any $g$, and the chain rule, we can then derive the useful projection identity

$$\int_{NP} f = -\frac{1}{2} \int_{abcd} \delta_{abcd} \int^C f \ . \tag{18}$$

Here $\int^C f$ denotes the indefinite integral of the arbitrary function $f(G_{Bab}, G_{Bcd}, C)$ in the variable $C$. The integral on the left hand side is now restricted to the nonplanar sector. For $f$ the integrand of our formula eq. (16), we have

$$\begin{aligned}\int^C f &= -\frac{C}{G_{Bab}G_{Bcd}} + \frac{4}{C}\ln\left(1 - \frac{C^2}{4G_{Bab}G_{Bcd}}\right) \\ &\quad + \frac{4}{\sqrt{G_{Bab}G_{Bcd}}}\,\text{arctanh}\left(\frac{1}{2}\frac{C}{\sqrt{G_{Bab}G_{Bcd}}}\right)\end{aligned} \tag{19}$$

We use symmetry to set $\delta_{abcd} = 4\,\delta(b-d)$, and the global invariance for setting $b \equiv 0$. This leaves us with three two-parameter integrals, of which the first one is elementary, while the second and third one can be reduced to known standard integrals by the substitutions

$$y = \frac{c(1-a)}{a(1-c)} \quad \text{and} \quad y^2 = \frac{c(1-a)}{a(1-c)} \ ,$$

respectively. The final result is

$$I_{NP} = 12\zeta(3) - 8\zeta(2) \ .$$

A similar but simpler calculation for the planar sector yields

$$I_P = 4\zeta(2) - 4 \ .$$

Let us mention in passing that the same integral appears in the calculation of the 3 - loop $\beta$ - function for $\phi^4$ - theory. This permits an easy check of the above calculation against a Feynman diagram calculation.

In diagrammatic terms, the integral $I(D)$ corresponds to a weighted sum of the two scalar 3-loop vertex diagrams depicted in fig. 2, calculated at zero external momentum, with massive propagators along the loop, and massless propagator insertions.

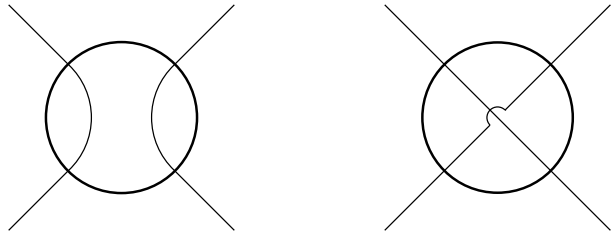

Figure 2:      $D_P$                 $D_{NP}$

It can be verified that, with an appropriate normalization, the relation

$$\Gamma(6 - \frac{3}{2}D)\,(I_{sing}(D) + I_P + I_{NP}) = 16D_P + 8D_{NP}$$

indeed holds true for the singular parts of the $\frac{1}{\epsilon}$ - expansions.

## 6. Discussion

We have given further examples for the efficiency of the worldline path integral approach to the Bern-Kosower formalism, and for the concept of multiloop worldline Green functions [7, 8, 26]. As a rule, it turns out to be advantageous to absorb a maximum amount of information into the worldline propagators themselves. A method of calculating three-loop parameter integrals has been developed which appears to be adequate for the calculation of three-loop renormalization group functions in those models for which the worldline method applies. At present this includes $\phi^3$-theory [7], $\phi^4$-theory [9], scalar [8, 9] and spinor [8] QED, and the Yukawa model [27].

# References

[1] Z. Bern, D. A. Kosower, Phys. Rev. Lett. **66** (1991) 1669;
    Z. Bern, D. A. Kosower, Nucl. Phys. **B379** (1992) 451.

[2] Z. Bern, L. Dixon, D. A. Kosower, Phys. Rev. Lett. **70** (1993) 2677.

[3] Z. Bern, D. C. Dunbar, T. Shimada, Phys. Lett. **B312** (1993) 277;
    D.C. Dunbar, P. S. Norridge, Nucl. Phys. **B433** (1995) 181.

[4] M. J. Strassler, Nucl. Phys. **B385** (1992) 145.

[5] L. Brink, P. Di Vecchia, P. Howe, Nucl.Phys. **B118** (1977) 76.

[6] A. M. Polyakov, *Gauge Fields and Strings*, Harwood 1987.

[7] M. G. Schmidt, C. Schubert, Phys. Lett. **B331** (1994) 69.

[8] M.G. Schmidt, C. Schubert, Phys. Rev. **D53** (1996) 2150.

[9] K. Daikouji, M. Shino, Y. Sumino, Phys. Rev. **D 53** (1996) 4598.

[10] M. G. Schmidt, C. Schubert, Phys. Lett. **B318** (1993) 438.

[11] D. Cangemi, E. D'Hoker, G. Dunne, Phys. Rev. **D51** (1995) 2513;
     R. Zh. Shaisultanov (hep-th/9512142).

[12] M. Reuter, M.G. Schmidt, C. Schubert, HD-THEP-96/17.

[13] S.L. Adler, C. Schubert, IASSNS-HEP-96/37 (hep-th/9605035).

[14] D. Fliegner, M.G. Schmidt, C. Schubert (work in progress).

[15] S.L. Adler, Ann. Phys. **67** (1971) 599.

[16] M. Mentzel, D. Berg, G. Wunner, Phys. Rev. **D 50** (1994) 1125.

[17] G. Wunner, R. Sang, D. Berg, Astrophys. J. **455** (1995) L51.

[18] C. Wilke, G. Wunner (hep-th/9605056).

[19] D.G.C. McKeon, T.N. Sherry, Mod. Phys. Lett. **A9** (1994) 2167.

[20] R.J. Stoneham, J. Phys. **A 12** (1979) 2187.

[21] V.N. Baier, A.I. Milstein, R. Zh. Shaisultanov (hep-th/9604028).

[22] W. Dittrich and M. Reuter, *Effective Lagrangians in Quantum Electrodynamics*, Springer 1985.

[23] J.L. Rosner, Ann. Phys. **44** (1967) 11; E. de Rafael and J.L. Rosner, Ann. Phys. **82** (1974) 369.

[24] D.J. Broadhurst, R. Delbourgo, D. Kreimer, Phys. Lett. **B366** (1996) 421.

[25] H.E. Brandt, Ph.D. thesis, University of Washington, 1970 (unpublished).

[26] K. Roland, H. Sato, NBI-HE-96-19 (hep-th/9604152).

[27] M. Móndragon, L. Nellen, M.G. Schmidt, C. Schubert, Phys. Lett. **B351** (1995) 200; E. D'Hoker, D. G. Gagné, UCLA-95-TEP-22 (hep-th/9508131).